\documentclass[journal=jacsat,manuscript=article]{achemso}
\usepackage{chemformula} 
\usepackage[T1]{fontenc} 

\usepackage{graphicx,epsfig,color,dcolumn,amssymb,amsmath}
\usepackage{epsfig}
\usepackage{epstopdf}
\usepackage{booktabs}
\usepackage{float}
\usepackage{placeins}
\usepackage{subcaption}
\usepackage{textgreek}
\usepackage{adjustbox}
\usepackage{color}
\usepackage{multirow}
\usepackage[version=4]{mhchem}
\RequirePackage{fix-cm} 

\usepackage[
colorlinks=true,
urlcolor=blue,
linkcolor=blue,
citecolor=blue
]{hyperref}

\author{Bill C. Oyomo}
\email{billc.oyomo@gmail.com}
\affiliation[The Technical University of Kenya]
{Materials Modeling Group, Department of Physics, Earth and Environmental Sciences, The Technical University of Kenya, P.O. Box 52428 - 00200, Nairobi, Kenya}

\author{Leah W. Mungai}
\affiliation[The Technical University of Kenya]
{Materials Modeling Group, Department of Physics, Earth and Environmental Sciences, The Technical University of Kenya, P.O. Box 52428 - 00200, Nairobi, Kenya}

\author{Geoffrey Arusei}
\affiliation[Kabianga University]
{Department of Physics, Kabianga University, P.O. Box 2030, 20200, Kericho, Kenya}

\author{Michael Atambo}
\affiliation[The Technical University of Kenya]
{Materials Modeling Group, Department of Physics, Earth and Environmental Sciences, The Technical University of Kenya, P.O. Box 52428 - 00200, Nairobi, Kenya}

\author{Mirriam Chepkoech}
\affiliation[The Technical University of Kenya]
{Materials Modeling Group, Department of Physics, Earth and Environmental Sciences, The Technical University of Kenya, P.O. Box 52428 - 00200, Nairobi, Kenya}

\author{Nicholas Makau}
\affiliation[University of Eldoret]
{Department of Physics, University of Eldoret, P.O. Box 1125, 30100, Eldoret, Kenya}

\author{G O Amolo}
\email{george.amolo@tukenya.ac.ke}
\affiliation[The Technical University of Kenya]
{Materials Modeling Group, Department of Physics, Earth and Environmental Sciences, The Technical University of Kenya, P.O. Box 52428 - 00200, Nairobi, Kenya}

\title{Thermoelastic Properties of the Ti$_2$AlC MAX Phase: An \textit{ab initio} study.}

\begin{document}

\begin{abstract}
MAX phases are used on an industrial scale in the transportation, armour, and
furnace development sectors, among others. However, data on the dynamical properties of these materials under varying temperature and pressure conditions are rare or unavailable. This study reports on the dynamical properties of the elastic constants of Ti$_2$AlC, under these conditions, obtained from first-principles calculations. Both static and dynamical results are presented and discussed. The dynamical results show that the elastic moduli are degraded; specifically, the bulk and shear moduli show a reduction ranging from 15 to 29\% and 13 to 31\%, respectively, between pressures of 10 - 30 GPa and in the temperature range of 300 - 1200 K. 
The reduction in these moduli is likely caused by anharmonic lattice effects that lead to thermal-induced softening, particularly in the high-pressure and temperature range. The lattice parameters of this material under the conditions of study did not vary significantly. Such data is useful as part of decision support that can inform applications as well as the limitations of use.
\end{abstract} 
\maketitle

\section{Introduction}
The MAX phases have been known and studied for decades, as captured in a recent review article by Dahlqvist, Barsoum, and Rosen~\cite{dahlqvist2024max}, which provides a rich source of their historical, current, and projected research development aspects as well as information ranging from traditional experimental synthesis of these materials to data obtained by methods independent from simulation. There are more than 300 MAX phases documented to be in existence. Despite this long period, their chemical diversity~\cite{sokol2019chemical} and their wide range of applications, such as solid solutions~\cite{ali2020recently} to attain new desired properties, potential superconductors~\cite{karacaprediction} and the hard materials industry~\cite{ali2021physical,ali2020recently,ali2021newly}, among others, have maintained a constant research interest in them to date. These phases having the general formula $M_{n+1}AX_{n}$ (n=1,2,3) are a class of layered ternary carbides and nitrides in which M, A, and X represent an early transition metal, an element of groups III-VI, and carbon or nitrogen, respectively~\cite{arusei2024elastic,barsoum2000mn+,hou2022fabrication_Ti2AlC,mo_MAX2012electronic}. Their layered crystal structure, which alternates between strong M-X bonds and weaker M-A bonds, contributes to their remarkable mechanical, chemical, and thermal stability under extreme conditions~\cite{arusei2024elastic, barsoum2000mn+,eklund2010mn+, barsoum2004mechanical}. MAX phase materials have found various applications in the industry that involve varying pressure and temperature conditions, such as corrosion-resistant materials~\cite{guo2024recent}, in heat exchangers and kiln furnaces, among others. 

The Ti$_2$AlC MAX phase is considered to be representative of these materials, as it has been extensively used in various applications.
There are reports of experimental high-temperature structural studies of bulk Ti$_2$AlC using neutron diffraction techniques~\cite{lane2013high} and phase content investigations~\cite{Pang_2010} in the 50 – 1200$^0$C and room temperature to 1550$^0$C, regimes, respectively, all of which show that this material maintains its geometrical structure under these conditions. Investigations of thin films~\cite{garkas2010synthesis} of the same material suggest that there is diffusion of Al out of the Ti$_2$AlC structure, resulting in the formation of other structures such as TiC and Ti$_3$AlC$_2$, implying the occurrence of phase transitions. The case of thin films is useful in accounting for surface phenomena such as the corrosion resistance application of Ti$_2$AlC through the formation of a protective alumina oxide layer following heating and hence oxidation in air~\cite{HAFTANI201651}. However, supporting simulations are lacking that would account for the application of both high pressure and temperature associated with the application of  Ti$_2$AlC. Finite temperature~\cite{DUONG2013296} simulation of  Ti$_2$AlC has shown the expected decrease in the elastic constants with rising temperature. However, this study~\cite{DUONG2013296} does not consider the effect of pressure, whose application to bulk crystals, without heating, would normally result in an increase in the elastic constants. The simultaneous consideration of both pressure and temperature would be expected in situations where such materials find applications. While it is expected that a combination of the two outcomes, as observed from their independent applications to crystals, will emerge, predicting the response of the material beyond a certain pressure and temperature is not obvious. 

This work has employed both high pressure and temperature to study the elastic constants’ response under these dynamic conditions, which has provided independent performance findings from experiment, some of which report varying observations as highlighted in the paragraph immediately above. It is expected that the matter of phase stability will emerge. First principle approaches are employed using the Quantum Espresso (QE)~\cite{QEscandolo2005first, giannozzi2009quantum} code as well as efficient processing tools such as the C$_{ij}$ code~\cite{luo2021cij} to evaluate thermoelastic properties and Snakemake~\cite{koster2012snakemake} for the management of workflows and efficient processing of the data.

\section{Methodology}
This study employed first-principle calculations within the framework of Density Functional Theory (DFT)~\cite{sholl2011density, kohn1965self, Hohenberg-64} as implemented in the Quantum Espresso code~\cite{giannozzi2009quantum,QEscandolo2005first}, with 60 Ry used as the energy cutoff (ecutwfc) for the plane wave basis set. Electron-ion interactions were treated using ultrasoft pseudopotentials. The exchange and correlation functional potentials were estimated using the local density approximation in the form of Perdew-Zunger (PZ) functionals~\cite{PZPhysRevB.23.5048}. A 12 x 12 x 2 Monkhorst-Pack k-point mesh was used to sample the first Brillouin zone. Structures at higher pressures were generated by uniformly scaling the ambient pressure structure followed by relaxation at fixed volumes. The static elastic tensor was then derived from the stress-strain relationships denoted by equation~\eqref{stress-strain}, applying small strains of ±0.5\%. 
\begin{equation}\label{stress-strain}
    \sigma_{ij} = C_{ijkl}\epsilon_{jk},
\end{equation}
where $C_{ijkl}$ is the $4^{th}$ order elastic tensor, while $\sigma_{ij}$ and $\epsilon_{kl}$ represent the $2^{nd}$ order stress and strain tensor, respectively~\cite{da2013ab}. 

A 2 x 2 x 2 q-point mesh was used in the QE code to obtain the static energies and phonon frequencies for the phonon calculations at different volumes or pressures.
The Snakemake workflow management system~\cite{koster2012snakemake} was used to automate the whole process of generating input files, submitting jobs, and collecting results for both static elasticity and phonon calculations at different pressures. The post-processed outputs for static elasticity and phonon calculations from QE were then collected and used as inputs in the Cij code~\cite{luo2021cij} to calculate the elastic tensor at elevated temperatures. The elastic coefficients or elements of the isothermal elastic tensor are expressed as strain derivatives of the Helmholtz free energy~\cite{luo2021cij,wu2011quasiharmonic,wang2024ab}:
\begin{equation}\label{elastic_tensor}
    c_{ijkl}^T = \frac{1}{V}\left( \frac{\delta^2F}{\delta e_{ij}\delta e_{kl}} \right) + \frac{1}{2}P(2\delta_{ij}\delta_{kl} - \delta_{il}\delta_{jk} - \delta_{ik}\delta_{jl})
\end{equation}
where, $e_{ij}$, P and F represent small strains, pressure, and the Helmholtz free energy, respectively~\cite{wu2011quasiharmonic,wang2024ab}. 
The Helmholtz free energy was computed based on the quasiharmonic approximation (QHA)~\cite{qin2019qha} as described by the following equation~\cite{wu2011quasiharmonic,wang2024ab}:
\begin{equation}
    F(e,V,T) = U^{st}(e,V) + \sum_qm \frac{1}{2}\hbar \omega_{qm}(e,V) + k_{b}T\sum_{qm}ln \left[ 1 - exp\left( \frac{-\hbar \omega_{qm}(e,V)}{k_{B}T} \right) \right]
\end{equation}
where, $U^{st}(e,V)$, q, V, $k_B$, T, and $\omega_{qm}$ represent the static total energy, phonon wave vector, equilibrium volume, Boltzmann constant, temperature, and the phonon frequencies corresponding to the $m^{th}$ mode, respectively.

The density-functional
perturbation theory (DFPT) method was employed using the phonopy code~\cite{togo2015first,togo2013phonopy} to perform the phonon calculations to obtain the harmonic phonon dispersions and determine the dynamical stability at different pressure values. The inter-atomic force constants (IFCs) and the dynamical matrix were calculated using a 2 x 2 x 2 supercell with a 4 x 4 x 2 mesh and a total energy convergence threshold of 1 x 10$^{-8}$ eV. We performed molecular dynamics (MD) simulations on a 2 x 2 x 2 supercell using the VASP code~\cite{kresse1996efficient,kresse2001vasp,kresse2007vasp} at constant volume (V) and temperature (T $=$ 1200 K) controlled by the Nosé dynamics ~\cite{hoover1985canonical}. To obtain the anharmonic phonon dispersion from molecular dynamics,  we used the normal-mode-decomposition technique as implemented in the Dynaphopy code~\cite{carreras2017dynaphopy}
\section{Results and Discussion}
\subsection{Structural Properties}
\begin{figure}[!htbp]
    \centering
    \begin{subfigure}[b]{0.50\textwidth} 
        \includegraphics[width=\linewidth]{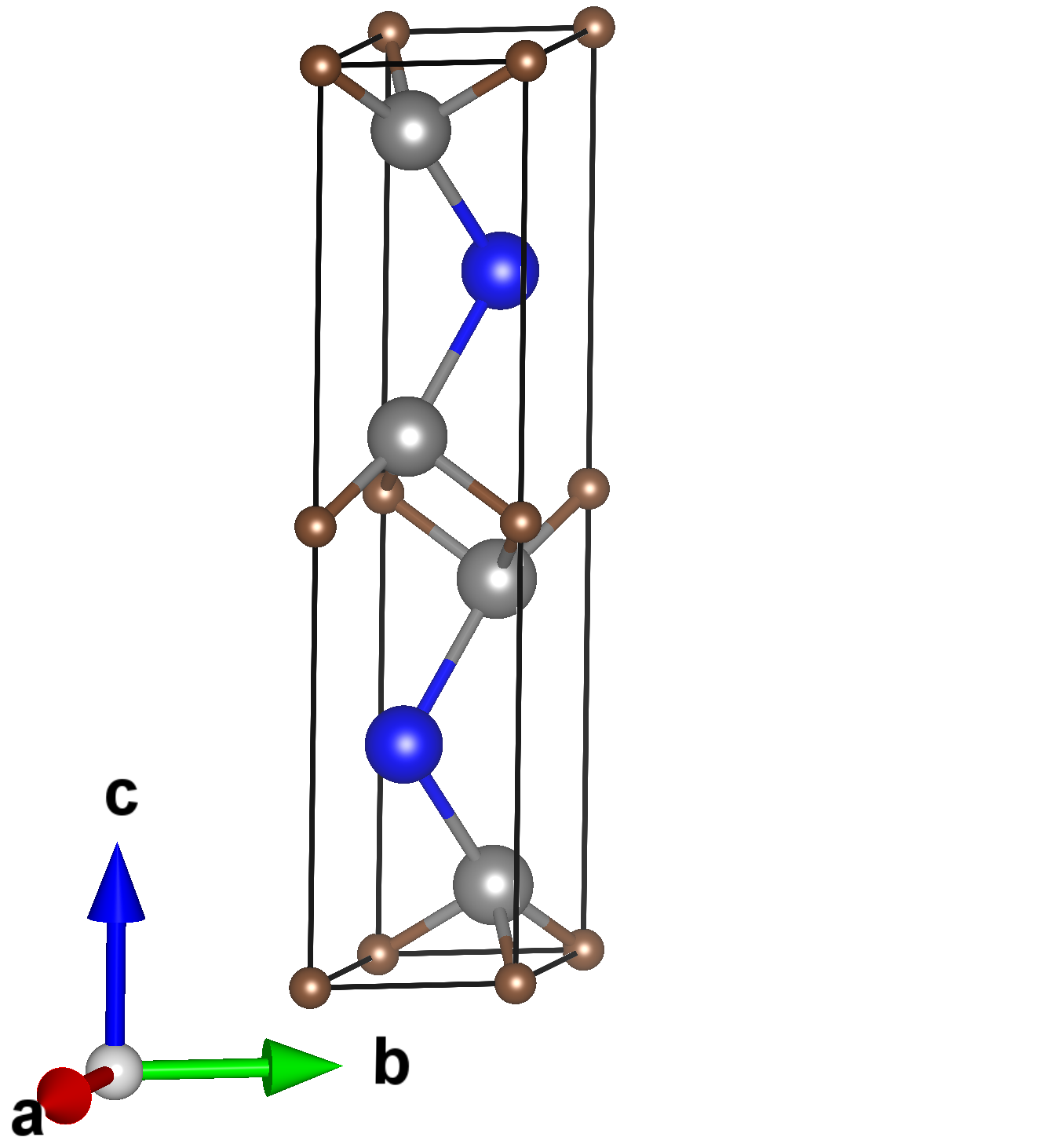}
        \caption{Optimized unit cell}
    \end{subfigure}
    \hspace{0.1cm} 
    \begin{subfigure}[b]{0.35\textwidth} 
        \includegraphics[width=\linewidth]{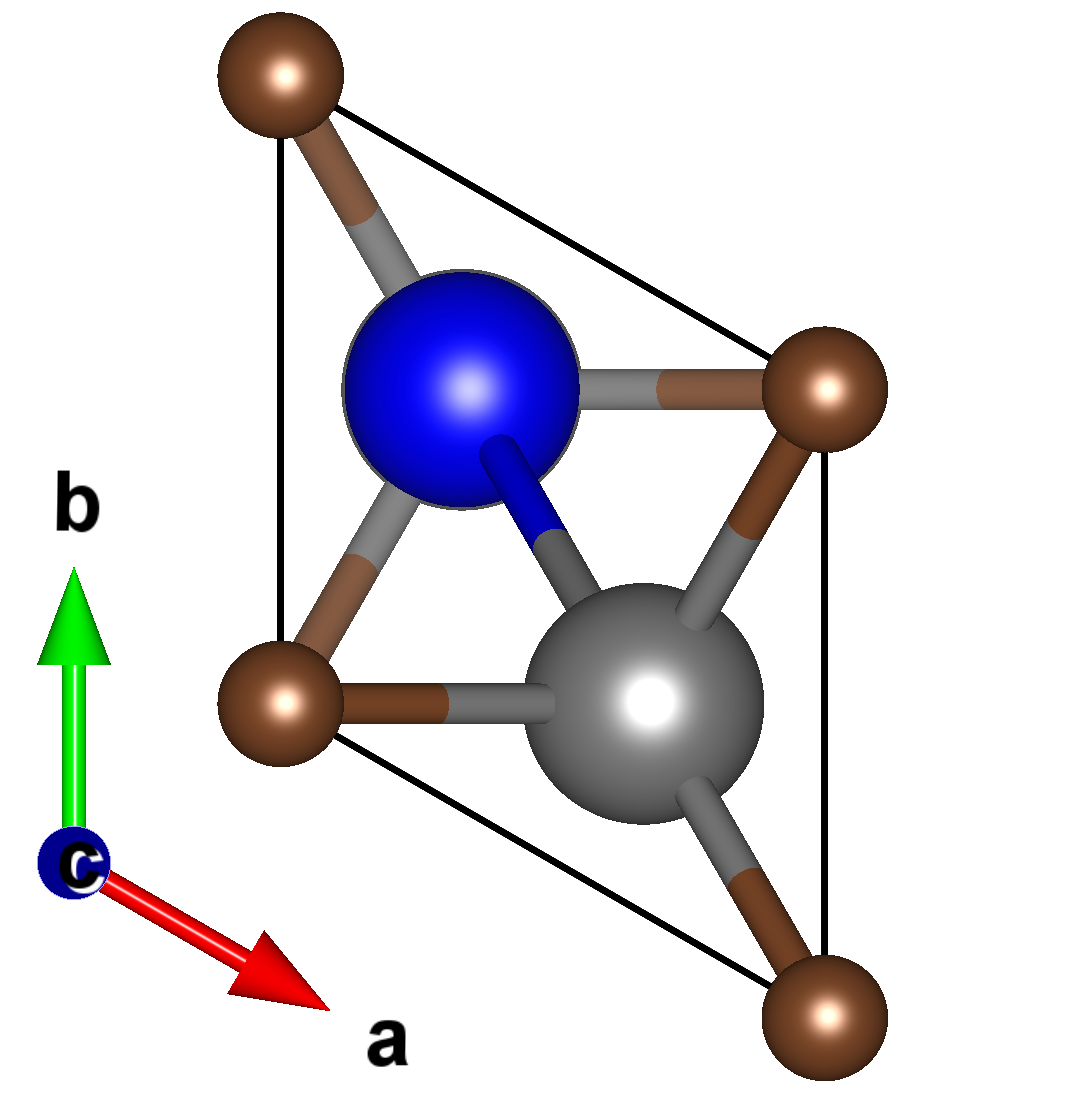}
        \caption{Basal plane showing the top view of the unit cell}
    \end{subfigure}
    \caption{The crystal structure of Ti$_2$AlC; (a) Optimized unit cell and (b) Basal plane showing the top view of the unit cell. The blue, grey, and brown balls represent Al, Ti, and C atoms, respectively.}
    \label{fig:Ti2AlC}
\end{figure}

Figure~\ref{fig:Ti2AlC} shows the hexagonal crystal structure of Ti$_2$AlC ($P6_3/mmc$ space group) with 8 atoms per unit cell. In this structure, there are four titanium ($Ti$) atoms occupying the 4f Wyckoff positions in which their z-coordinates are not fixed and can vary within the unit cell, two aluminium ($Al$) atoms located at the 2d positions, and two carbon ($C$) atoms residing at the 2a positions, both constrained by the symmetry~\cite{lin2006microstructural_Ti2AlC,mo_MAX2012electronic}. Each Ti atom is bonded to three equivalent Al atoms and three equivalent C atoms, forming a trigonal bipyramidal coordination (3-coordinate geometry). A previous study by Arusei and co-authors explored the elastic and thermal properties of several MAX phases using density functional theory, revealing that the melting temperatures of these materials range from 1100 to 1700 K, thus indicating their potential for high-temperature applications~\cite{arusei2024elastic}. Further, although the study considered the static elastic properties of 7 MAX phases, a focus on Ti$_2$AlC and Ti$_2$CdC, demonstrated volume reduction with increasing pressure, which is critical for understanding the behaviour of these materials under dynamic conditions.

\subsection{Structural Response to Pressure}
To understand the structural behavior of Ti$_2$AlC under pressure, we optimized the unit-cell parameters (\(a\), \(c\)) and the atomic positions at fixed volumes across a 0–37 GPa pressure range. The pressure dependence of lattice parameters, axial ratio (\(c/a\)), and unit cell volume reveals anisotropic compressibility, with the \(c\)-axis (Ti–Al bonds) being more compressible than the \(a\)-axis (Ti–C bonds), as demonstrated in Fig. S1 (see section II in the supplementary information for details). The volume data were fitted to the third-order Birch-Murnaghan equation of state as shown in Fig. S2 (supplementary information), yielding a bulk modulus of 152 GPa, an equilibrium volume of 106.38 \AA$^3$, and a pressure derivative of 4.38 (see the supplementary information, Table S1).

\subsection{Static values of the elastic constants, C\textsubscript{ij}s.}
Figure 2, on the left-hand panel, shows the values of the static elastic coefficients of Ti$_2$AlC as a function of pressure. An inspection of the lattice constants of this material before and after being subjected to the pressure range under investigation reveals no change in phase. Ti$_2$AlC has five independent elastic constants, i.e., $C_{11}$, $C_{12}$, $C_{13}$, $C_{33}$, and, $C_{44}$ based on its hexagonal structure. All values of $C_{ij}$ increase monotonically with rising pressure. $C_{11}$ and $C_{33}$ have higher values ranging, approximately, from 240 to 400 GPa and 180 to 380 GPa, respectively. The high values of $C_{11}$ and $C_{33}$ in Ti$_2$AlC under pressure stem from its layered structure and bonding. Strong covalent Ti-C bonds in the basal plane contribute to high stiffness along the a-axis (($C_{11}$): 240-400 GPa), while slightly weaker Ti-Al bonds between layers influence the c-axis (($C_{33}$): 180-380 GPa). As pressure increases, both types of bonds stiffen, resulting in a monotonic rise in elastic constants and an anisotropic response characteristic of MAX phases. $C_{12}$, $C_{13}$ and $C_{44}$ have lower values ranging approximately from 30-107 GPa, 15-115 GPa, and 81-149 GPa, respectively. The effect of increased pressure is to bring the atoms closer, making the material stiffer, which would explain the increase in the C\textsubscript{ij}s with increasing pressure.
The right-hand panel on this figure shows the variation of the same constants with pressure at 0 K obtained from thermoelastic calculations. 
The increase in $C_{11}$, $C_{12}$, and $C_{44}$ with pressure appears only modest, with $C_{11}$ ranging between 330 and 360 GPa, while $C_{12}$ ranges between 60 and 110 GPa and $C_{44}$ varies between 125 and 149 GPa for pressures of between 0 and 40 GPa. On the contrary $C_{33}$ and $C_{13}$ show significant increases with pressure ranging between 300-630 GPa and 70-198 GPa. $C_{33}$ clearly shows the largest increase over the pressure range of 0-35 GPa while $C_{12}$, $C_{44}$, and $C_{11}$ appear insensitive to thermoelastic considerations.
 
\begin{figure}[!thbp]
    \centering
    \begin{subfigure}[b]{0.49\textwidth} 
        \includegraphics[width=\linewidth]{Figures/Cij-Presure_Static.eps}
    \end{subfigure}
    \begin{subfigure}[b]{0.49\textwidth} 
        \includegraphics[width=\linewidth]{Figures/C_ij_0K.eps}
    \end{subfigure}
    \caption{The $C_{ij}$'s of Ti$_2$AlC versus pressure under static calculations (LHS panel) and from thermoelastic considerations (RHS panel), at 0 K.}
    \label{fig:pressure-Cij}
\end{figure}

The mechanical stability of Ti$_2$AlC was assessed using the generalized Born-Huang criteria for hexagonal crystals~\cite{mouhat2014necessary}, given its \(P6_3/mmc\) structure. All five elastic constants (\(C_{11}, C_{12}, C_{13}, C_{33}, C_{44}\)) satisfy the conditions \(C_{11} > |C_{12}|\), \(C_{44} > 0\), and \(2C_{13}^2 < C_{33}(C_{11} + C_{12})\) across 0–35 GPa, confirming stability under pressure. Detailed calculations are provided in the Supplementary Information, Section III, Table S2.

In addition, we calculated the bulk \& shear moduli, Young’s modulus (\(E\)), Poisson’s ratio (\(\nu\)), and Pugh’s ratio (\(B/G\)) to further characterize Ti$_2$AlC’s mechanical behaviour under compression. Both bulk and shear moduli, as well as Young’s modulus, increase with pressure, indicating enhanced stiffness. The Poisson’s ratio also rises from 0.014 to 0.225, suggesting a shift toward ductility, though remaining brittle (\(\nu < 0.26\)). Pugh’s ratio (\(B/G\)) also increases but stays below the ductility threshold of 1.75 (detailed discussion in the supplementary information, Section III, Table S2).
\FloatBarrier
\begin{figure}[!htbp]
    \centering
    \begin{subfigure}[b]{0.49\textwidth} 
        \includegraphics[width=\linewidth]{Figures/C11.eps}
        \caption{}
    \end{subfigure}
    \begin{subfigure}[b]{0.49\textwidth} 
        \includegraphics[width=\linewidth]{Figures/C12.eps}
        \caption{}
    \end{subfigure}
    \begin{subfigure}[b]{0.49\textwidth} 
        \includegraphics[width=\linewidth]{Figures/C13.eps}
        \caption{}
    \end{subfigure}
    \begin{subfigure}[b]{0.49\textwidth} 
        \includegraphics[width=\linewidth]{Figures/C33.eps}
        \caption{}
    \end{subfigure}
    \begin{subfigure}[b]{0.49\textwidth} 
        \includegraphics[width=\linewidth]{Figures/C44.eps}
        \caption{}
    \end{subfigure}
    \caption{The dynamical elastic constants, $C_{ij}$'s, of Ti$_2$AlC versus pressure at various temperatures for (a) $C_{11}$, (b) $C_{12}$, (c) $C_{13}$, (d) $C_{33}$ and (e) $C_{44}$.}
    \label{fig:PxT-Cij}
\end{figure}

\setlength{\tabcolsep}{4pt} 

\subsection{Elastic properties under dynamic conditions}
In Figure 3, the behaviour of the elastic constants under varying conditions of temperature and pressure is shown.
It is observed that all 5 elastic constants are showing softening with the increase in pressure across the temperature profiles studied and in the 0–1100 K range. Only selected profiles are shown in the current graphs. The upper-temperature limit of 1200 K is close to the predicted melting point of the material. These calculations assume that the materials under investigation are pristine. Variations may be expected if the calculations include simple or extended defects as would occur in nature. However, this was not within the scope of the current study. It is observed that $C_{11}$, $C_{12}$ and $C_{44}$ show similar trends under varying conditions of rising temperature and pressure, leading to a reduction in the values of the elastic constants. $C_{13}$ and $C_{33}$, while showing decreasing values of the elastic constants with temperature and pressure, the trends are different from the other 3 $C_{ij}$s. At pressures above 20 GPa, the rate of increase in $C_{13}$ and $C_{33}$ appears to be much faster than at the lower temperatures. Introduction of  thermal energy into the material has the effect of increasing the vibrational amplitudes about the mean atomic positions, making the system less stiff. It is expected that the pronounced decrease in the C\textsubscript{ij}s observed towards the melting point is when the atoms have moved away from their mean atomic positions and some disorder has emerged. 

The combined effect of the application of increasing temperature and pressure is exhibited in a complex way. For $C_{11}$, $C_{12}$ and $C_{44}$, the increase in the values depends on the temperature, displaying a linear increase but with gentle curvature from 0 to 1050 K and 0 to 1200 K profiles for $C_{11}$ and $C_{12}$, respectively, while remaining almost linear for $C_{44}$, in a similar temperature range. All these are observed between 0 and 15 GPa. Above this pressure range, an enhanced decrease is noted for all three elastic constants.

\begin{figure}[!htbp]
    \centering
    \begin{subfigure}{0.49\linewidth}
        \centering
        \includegraphics[width=\linewidth]{Figures/Bulk_modulus.eps}
        \caption{}
        \label{fig:Bulk_modulus-Temp}
    \end{subfigure}
    \hfill
    \begin{subfigure}{0.49\linewidth}
        \centering
        \includegraphics[width=\linewidth]{Figures/Shear-Modulus.eps}
        \caption{}
        \label{fig:Shear_modulus-Temp}
    \end{subfigure}
    \caption{The dynamical elastic moduli of Ti$_2$AlC at various pressures and temperatures: (a) Bulk modulus and (b) Shear modulus}
    \label{fig:Bulk-Shear-Modulus_Temp}
\end{figure}

Figure~\ref{fig:Bulk-Shear-Modulus_Temp} shows the behaviour of the bulk and shear moduli under dynamic conditions of pressure and temperature. There are clear observations of the degradation of the moduli as a function of temperature at selected pressures, as indicated in Table~\ref{tab:change_in_K_G}. These results suggest that in the temperature range under consideration, the bulk modulus will experience a variation of up to 15\%, 24\%, and 29\% at pressures of 10, 20, and 30 GPa, respectively. Similarly, the shear modulus will vary by 13\%, 20\%, and 31\% at the same pressures, respectively. The results show that for applications at other fixed pressures and varying temperatures, there are notable changes in both the bulk and shear moduli. This suggests that the dynamic conditions have an overall effect on the elastic constants of Ti$_2$AlC as well as its bulk and shear modulus. It is observed that the trend in the degradation of the bulk and shear moduli has a similar form as $C_{11}$, $C_{12}$ and $C_{44}$. This is more clearly depicted in the shear modulus.

\begin{table}
    \centering
    \caption{Changes in the dynamic bulk and shear moduli between 300 and 1200 K.}
    \begin{tabular}{ccc}
    \hline
    \hline
        Pressure & \% change of bulk modulus, \% change of Shear modulus \\
    \hline
    \hline
       10  & 15.48 & 13.23\\
       20  & 23.97 & 20.41\\
       30  & 29.20 & 31.18\\
    \hline
    \hline
    \end{tabular}
    \label{tab:change_in_K_G}
\end{table}

The pronounced softening of the shear modulus beyond 15 GPa at 1200 K, as shown in Figure ~\ref{fig:Shear_modulus-Temp}, could suggest underlying structural or dynamical changes, potentially indicative of melting or amorphization, as noted in the study by Singh et al. on AuI \cite{singh2022high}. In their work, pressure-driven amorphization in AuI at 7 GPa was linked to phonon softening and violation of the Born stability criteria, signaling mechanical and dynamical instabilities~\cite{singh2022high}. In contrast, our calculations for Ti$_2$AlC, the estimated elastic constants, indicate that the Born stability criteria for the hexagonal lattice are satisfied across the studied pressure range (details in section III, Table S2 in the supplementary information), confirming mechanical stability. To assess potential structural compromise, we analyzed the bond angles and bond lengths as a function of pressure shown in Fig. S3 (see the supplementary information). These results indicate comparable values of lattice parameters with the high-temperature neutron diffraction experimental work reported elsewhere by Lane and coworkers \cite{lane2013high}. From the current work, we observe  systematic and modest changes that reflect anisotropic compression of the layered MAX phase, with greater compressibility along the c-axis (Ti-Al bonds) than in the a-b plane (Ti-C, Ti-Ti bonds), and no abrupt distortions around 15 GPa. This behaviour indicates that the hexagonal phase persists across the study pressure range up to $\approx37$ GPa, ruling out amorphization or melting, the latter being unlikely as 1200 K is well below the reported melting point of Ti$_2$AlC (1700 K) \cite{arusei2024elastic, pan2024exploring}. Therefore, we propose that the softening may be attributed to pressure-temperature-induced anharmonic lattice effects.

\begin{figure}[!htbp]
    \centering
    \begin{subfigure}{0.49\linewidth}
        \centering
        \includegraphics[width=\linewidth]{Figures/bm_Temp.eps}
        \caption{}
        \label{fig:bm-Temp}
    \end{subfigure}
    \hfill
    \begin{subfigure}{0.49\linewidth}
        \centering
        \includegraphics[width=\linewidth]{Figures/G_Temp.eps}
        \caption{}
        \label{fig:G-Temp}
    \end{subfigure}
    \caption{Temperature dependence of the elastic moduli of Ti$_2$AlC at various pressures: (a) Bulk modulus and (b) Shear modulus.}
    \label{fig:bn-G_Temp}
\end{figure}

Figure~\ref{fig:bn-G_Temp} shows the results of the bulk and shear moduli as a function of temperature for pressure values ranging from 0 to 35 GPa. The results of the bulk modulus on the left-hand-side panel not only show comparable values at 0 K but also a similar trend with the rise in the temperature by Duong et al.~\cite{DUONG2013296}. The experimental results of Radovic et al.~\cite{radovic2006elastic} and simulations by Duong et al.~\cite{DUONG2013296} are shown for the shear modulus profile as a way of comparison. For the case of works of Benitez and co-workers~\cite{benitez2020mechanical} where compression strength is studied as a function of rising temperature, these moduli decrease for a fixed pressure. It can be observed, in both cases, that the trends of the decrease of the shear modulus with rising temperature are similar to the current work near the 0 GPa profile, and the values are within a 10\% margin of error in reference to the experimental work, which would be considered an independent approach. An account for these differences is made as follows: the calculations in the current work assume a pristine crystal structure, while it is expected that the real material would not be defect-free. The simulation results of Duong et al.\cite{DUONG2013296} have not considered the simultaneous inclusion of the effects of pressure. 

\section{Phonon stability at high-pressure and temperature conditions}
To explore the dynamic stability of the Ti$_2$AlC MAX phase, we have calculated both the harmonic and anharmonic phonon dispersion at different pressures and 1200 K temperature, respectively. Figure~\ref{fig:Phonons_TP} shows the dispersion profiles for the pressure levels under study as well as the renormalized dispersion at 1200 K. In all pressure values, there is no negative phonon frequency, which suggests that the   Ti$_2$AlC is dynamically stable at high pressures. Additionally, the renormalized anharmonic phonon dispersion at 1200 K, shown as red lines in Figure~\ref{HRPD1200K}, also indicates that the material is dynamically stable at that temperature due to the absence of imaginary frequencies. 
Although the structure remains dynamically stable at the elevated pressures and temperature, the effect of temperature on the phonon frequencies is very noticeable. For instance, the renormalized phonon dispersion at the 1200 K temperature shows a general upward shift in the phonon frequencies except for one optical phonon mode close to 20 THz that shifts downwards near $\Gamma$.

\FloatBarrier
\begin{figure}[!htbp]
    \centering
    \begin{subfigure}[b]{0.49\textwidth} 
        \includegraphics[width=\linewidth]{Figures/Phonon.bands.0GPa.eps}
        \caption{}
    \end{subfigure}
    \begin{subfigure}[b]{0.49\textwidth} 
        \includegraphics[width=\linewidth]{Figures/Phonon.bands.24GPa.eps}
        \caption{}
    \end{subfigure}
    \begin{subfigure}[b]{0.55\textwidth} 
        \includegraphics[width=\linewidth]{Figures/Harmonic_Renormalized_Phonon_Dispersion.eps}
        \caption{}
        \label{HRPD1200K}
    \end{subfigure}
    \caption{The phonon dispersions for Ti$_2$AlC at three different pressures: (a) 0 GPa, (b) 24 GPa, and (c) 35 GPa. The black and red lines represent temperatures of 0 K and 1200 K, respectively.}
    \label{fig:Phonons_TP}
\end{figure}

\newpage
\section{Conclusion}
Elastic constants of bulk Ti$_2$AlC have been studied under dynamic pressure and temperature conditions in the 0–1200 K range with pressures ranging from 0 to 35 GPa. These investigations have employed \textit{ab initio} approaches and postprocessing techniques in the study of this material. All  elastic constants have shown anisotropic reduction in their values under dynamic conditions compared to those obtained purely under the static case: The bulk and shear moduli show reductions of up to 30\% at temperatures close to 1200 K relative to room temperature of 300 K, suggesting limits on where these materials are likely to find applications. This decrease in moduli at elevated temperatures can be attributed to anharmonic lattice effects that manifest as thermal-induced softening, which results in the reduction of the material’s resistance to deformation under dynamic loading. Lack of negative frequencies further confirms the stability of the bulk Ti$_2$AlC MAX phase at elevated temperatures of up to 1200 K and pressures of up to 35 GPa considered in this study.

\begin{suppinfo}
 Supporting Information: Pressure-dependent lattice parameters, axial ratio (c/a), and unit cell volume; Birch–Murnaghan EOS fitting parameters; elastic constants and derived mechanical properties (bulk, shear, and Young’s moduli, Poisson’s ratio, Pugh’s ratio); bond lengths and bond angles under pressure; Figures S1–S3; Tables S1–S2.   
\end{suppinfo}

\begin{acknowledgement}
 The authors acknowledge the Centre for High-Performance Computing (CHPC, South Africa) for the computational resources allocated to the MATS862 account and access to the Kenya Education Network (KENET) GPU cluster for compute time.   
\end{acknowledgement}

\bibliography{references}

@book{sholl2011density,
  title={Density functional theory: a practical introduction},
  author={Sholl, David and Steckel, Janice A},
  year={2011},
  publisher={John Wiley \& Sons}
}

@article{Hohenberg-64,
title = {Inhomogeneous Electron Gas},
author = {Hohenberg, P. and Kohn, W.},
journal = {Phys. Rev.},
volume = {136},
issue = {3B},
pages = {B864--B871},
numpages = {0},
year = {1964},
month = {Nov},
publisher = {American Physical Society},
doi = {10.1103/PhysRev.136.B864},
url = {https://link.aps.org/doi/10.1103/PhysRev.136.B864}
}

@article{togo2015first,
  title={First principles phonon calculations in materials science},
  author={Togo, Atsushi and Tanaka, Isao},
  journal={Scripta Materialia},
  volume={108},
  pages={1--5},
  year={2015},
  publisher={Elsevier}
}

@article{kohn1965self,
  title={Self-consistent equations including exchange and correlation effects},
  author={Kohn, Walter and Sham, Lu Jeu},
  journal={Physical review},
  volume={140},
  number={4A},
  pages={A1133},
  year={1965},
  publisher={APS}
}

@misc{togo2013phonopy,
  title={phonopy manual},
  author={Togo, Atsushi},
  year={2013}
}

@article{mouhat2014necessary,
  title={Necessary and sufficient elastic stability conditions in various crystal systems},
  author={Mouhat, F{\'e}lix and Coudert, Fran{\c{c}}ois-Xavier},
  journal={Physical review B},
  volume={90},
  number={22},
  pages={224104},
  year={2014},
  publisher={APS}
}

@article{hoover1985canonical,
  title={Canonical dynamics: Equilibrium phase-space distributions},
  author={Hoover, William G},
  journal={Physical review A},
  volume={31},
  number={3},
  pages={1695},
  year={1985},
  publisher={APS}
}

@article{giannozzi2009quantum,
  title={QUANTUM ESPRESSO: a modular and open-source software project for quantum simulations of materials},
  author={Giannozzi, Paolo and Baroni, Stefano and Bonini, Nicola and Calandra, Matteo and Car, Roberto and Cavazzoni, Carlo and Ceresoli, Davide and Chiarotti, Guido L and Cococcioni, Matteo and Dabo, Ismaila and others},
  journal={Journal of physics: Condensed matter},
  volume={21},
  number={39},
  pages={395502},
  year={2009},
  publisher={IOP Publishing}
}

@article{QEscandolo2005first,
  title={First-principles codes for computational crystallography in the Quantum-ESPRESSO package},
  author={Scandolo, Sandro and Giannozzi, Paolo and Cavazzoni, Carlo and de Gironcoli, Stefano and Pasquarello, Alfredo and Baroni, Stefano},
  journal={Zeitschrift f{\"u}r Kristallographie-Crystalline Materials},
  volume={220},
  number={5-6},
  pages={574--579},
  year={2005},
  publisher={De Gruyter Oldenbourg}
}

@article{luo2021cij,
  title={cij: A Python code for quasiharmonic thermoelasticity},
  author={Luo, Chenxing and Deng, Xin and Wang, Wenzhong and Shukla, Gaurav and Wu, Zhongqing and Wentzcovitch, Renata M},
  journal={Computer Physics Communications},
  volume={267},
  pages={108067},
  year={2021},
  publisher={Elsevier}
}

@article{PZPhysRevB.23.5048,
  title = {Self-interaction correction to density-functional approximations for many-electron systems},
  author = {Perdew, J. P. and Zunger, Alex},
  journal = {Phys. Rev. B},
  volume = {23},
  issue = {10},
  pages = {5048--5079},
  numpages = {0},
  year = {1981},
  month = {May},
  publisher = {American Physical Society},
  doi = {10.1103/PhysRevB.23.5048},
  url = {https://link.aps.org/doi/10.1103/PhysRevB.23.5048}
}

@article{wu2011quasiharmonic,
  title={Quasiharmonic thermal elasticity of crystals: An analytical approach},
  author={Wu, Zhongqing and Wentzcovitch, Renata M},
  journal={Physical Review B—Condensed Matter and Materials Physics},
  volume={83},
  number={18},
  pages={184115},
  year={2011},
  publisher={APS}
}

@article{wang2024ab,
  title={Ab initio study of the stability and elasticity of brucite},
  author={Wang, Hongjin and Luo, Chenxing and Wentzcovitch, Renata M},
  journal={Physical Review B},
  volume={109},
  number={21},
  pages={214103},
  year={2024},
  publisher={APS}
}

@article{koster2012snakemake,
  title={Snakemake—a scalable bioinformatics workflow engine},
  author={K{\"o}ster, Johannes and Rahmann, Sven},
  journal={Bioinformatics},
  volume={28},
  number={19},
  pages={2520--2522},
  year={2012},
  publisher={Oxford University Press}
}

@article{qin2019qha,
  title={qha: A Python package for quasiharmonic free energy calculation for multi-configuration systems},
  author={Qin, Tian and Zhang, Qi and Wentzcovitch, Renata M and Umemoto, Koichiro},
  journal={Computer Physics Communications},
  volume={237},
  pages={199--207},
  year={2019},
  publisher={Elsevier}
}

@article{arusei2024elastic,
  title={Elastic and thermal properties of selected 211 MAX phases: A DFT study},
  author={Arusei, GK and Chepkoech, M and Amolo, GO and Makau, NW},
  journal={Computational Condensed Matter},
  volume={39},
  pages={e00891},
  year={2024},
  publisher={Elsevier}
}

@inproceedings{da2013ab,
  title={Ab initio elasticity workflow in the VLab science gateway},
  author={Da Silveira, Pedro RC and Gunathilake, Lahiru and Holiday, Alexander and Yuen, Dave A and Valdez, M N{\'u}{\~n}ez and Wentzcovitch, Renata M},
  booktitle={Proceedings of the Conference on Extreme Science and Engineering Discovery Environment: Gateway to Discovery},
  pages={1--8},
  year={2013}
}

@article{barsoum2000mn+,
  title={The MN+ 1AXN phases: A new class of solids: Thermodynamically stable nanolaminates},
  author={Barsoum, Michel W},
  journal={Progress in solid state chemistry},
  volume={28},
  number={1-4},
  pages={201--281},
  year={2000},
  publisher={Elsevier}
}

@article{barsoum2004mechanical,
  title={Mechanical properties of the MAX phases},
  author={Barsoum, MW and Radovic, M},
  journal={Encyclopedia of materials: science and technology},
  volume={160},
  number={s1},
  pages={1--16},
  year={2004},
  publisher={Elsevier Oxford}
}

@article{guo2024recent,
  title={Recent progress in synthesis of MAX phases and oxidation \& corrosion mechanism: a review},
  author={Guo, Min and Cao, Guoqin and Pan, Haoyu and Guo, Junhong and Chen, Chaoyang and Zhang, Baofeng and Hu, Junhua},
  journal={Materials Research Letters},
  volume={12},
  number={11},
  pages={765--796},
  year={2024},
  publisher={Taylor \& Francis}
}

@article{mo_MAX2012electronic,
  title={Electronic structure and optical conductivities of 20 MAX-phase compounds},
  author={Mo, Yuxiang and Rulis, Paul and Ching, WY},
  journal={Physical Review B—Condensed Matter and Materials Physics},
  volume={86},
  number={16},
  pages={165122},
  year={2012},
  publisher={APS}
}

@article{lin2006microstructural_Ti2AlC,
  title={Microstructural characterization of layered ternary Ti2AlC},
  author={Lin, ZJ and Zhuo, MJ and Zhou, YC and Li, MS and Wang, JY},
  journal={Acta materialia},
  volume={54},
  number={4},
  pages={1009--1015},
  year={2006},
  publisher={Elsevier}
}

@article{hou2022fabrication_Ti2AlC,
  title={Fabrication, microstructure and compressive properties of Ti2AlC/TiAl composite with a bioinspired laminated structure},
  author={Hou, Bo and Liu, Pei and Wang, Aiqin and Xie, Jingpei},
  journal={Vacuum},
  volume={201},
  pages={111124},
  year={2022},
  publisher={Elsevier}
}

@article{eklund2010mn+,
  title={The Mn+ 1AXn phases: Materials science and thin-film processing},
  author={Eklund, Per and Beckers, Manfred and Jansson, Ulf and H{\"o}gberg, Hans and Hultman, Lars},
  journal={Thin Solid Films},
  volume={518},
  number={8},
  pages={1851--1878},
  year={2010},
  publisher={Elsevier}
}

@article{singh2022high,
  title={High-pressure study of the aurophilic topological Dirac material AuI},
  author={Singh, Jaspreet and Sahoo, Sushree Sarita and Venkatakrishnan, Kanchana and Vaitheeswaran, Ganapathy and Errandonea, Daniel},
  journal={Journal of Alloys and Compounds},
  volume={928},
  pages={167178},
  year={2022},
  publisher={Elsevier}
}

@article{pan2024exploring,
  title={Exploring the structural, phonon dynamical, mechanical and thermodynamic properties of TM2AlC (TM= Ti, Zr and Hf) carbides},
  author={Pan, Yong and Yang, Zhijing and Zhang, Hui},
  journal={Diamond and Related Materials},
  volume={144},
  pages={110966},
  year={2024},
  publisher={Elsevier}
}

@article{sokol2019chemical,
  title={On the chemical diversity of the MAX phases},
  author={Sokol, Maxim and Natu, Varun and Kota, Sankalp and Barsoum, Michel W},
  journal={Trends in Chemistry},
  volume={1},
  number={2},
  pages={210--223},
  year={2019},
  publisher={Elsevier}
}

@article{dahlqvist2024max,
  title={MAX phases--Past, present, and future},
  author={Dahlqvist, Martin and Barsoum, Michel W and Rosen, Johanna},
  journal={Materials Today},
  volume={72},
  pages={1--24},
  year={2024},
  publisher={Elsevier}
}

@article{ali2021newly,
  title={Newly synthesized MAX phase Zr 2 SeC: DFT insights into physical properties towards possible applications},
  author={Ali, MA and Qureshi, Muhammad Waqas},
  journal={RSC advances},
  volume={11},
  number={28},
  pages={16892--16905},
  year={2021},
  publisher={Royal Society of Chemistry}
}

@article{ali2021physical,
  title={Physical properties of new MAX phase borides M2SB (M= Zr, Hf and Nb) in comparison with conventional MAX phase carbides M2SC (M= Zr, Hf and Nb): Comprehensive insights},
  author={Ali, MA and Hossain, MM and Uddin, MM and Hossain, MA and Islam, AKMA and Naqib, SH},
  journal={Journal of Materials Research and Technology},
  volume={11},
  pages={1000--1018},
  year={2021},
  publisher={Elsevier}
}

@misc{karacaprediction,
  title={Prediction of phonon-mediated superconductivity in new Ti-based M2AX phases. Sci Rep 2022; 12: 13198},
  author={Karaca, E and Byrne, PJP and Hasnip, PJ and Probert, MIJ}
}

@article{ali2020recently,
  title={Recently synthesized (Ti 1- x Mo x) 2 AlC (0≤ x≤ 0.20) solid solutions: deciphering the structural, electronic, mechanical and thermodynamic properties via ab initio simulations},
  author={Ali, MA and Naqib, SH},
  journal={RSC advances},
  volume={10},
  number={52},
  pages={31535--31546},
  year={2020},
  publisher={Royal Society of Chemistry}
}

@article{radovic2006elastic,
  title={On the elastic properties and mechanical damping of Ti3SiC2, Ti3GeC2, Ti3Si0. 5Al0. 5C2 and Ti2AlC in the 300--1573 K temperature range},
  author={Radovic, Miladin and Barsoum, MW and Ganguly, A and Zhen, T and Finkel, P and Kalidindi, SR and Lara-Curzio, Edgar},
  journal={Acta materialia},
  volume={54},
  number={10},
  pages={2757--2767},
  year={2006},
  publisher={Elsevier}
}

@article{DUONG2013296,
title = {First-principles calculations of finite-temperature elastic properties of Ti2AlX (X=C or N)},
journal = {Computational Materials Science},
volume = {79},
pages = {296-302},
year = {2013},
issn = {0927-0256},
doi = {https://doi.org/10.1016/j.commatsci.2013.06.033},
url = {https://www.sciencedirect.com/science/article/pii/S0927025613003662},
author = {Thien Chi Duong and Navdeep Singh and Raymundo Arróyave},
}

@article{Pang_2010,
doi = {10.1088/1742-6596/251/1/012025},
url = {https://dx.doi.org/10.1088/1742-6596/251/1/012025},
year = {2010},
month = {nov},
publisher = {},
volume = {251},
number = {1},
pages = {012025},
author = {Pang, W K and Low, I M and O'Connor, B H and Studer, A J and Peterson, V K and Sun, Z M and Palmquist, J -P},
title = {Comparison of thermal stability in MAX211 and 312 phases},
journal = {Journal of Physics: Conference Series},
}

@article{lane2013high,
  title={High-temperature neutron diffraction and first-principles study of temperature-dependent crystal structures and atomic vibrations in Ti3AlC2, Ti2AlC, and Ti5Al2C3},
  author={Lane, Nina J and Vogel, Sven C and Caspi, El'ad N and Barsoum, Michel W},
  journal={Journal of applied physics},
  volume={113},
  number={18},
  year={2013},
  publisher={AIP Publishing}
}

@article{benitez2020mechanical,
  title={Mechanical properties and microstructure evolution of Ti2AlC under compression in 25--1100° C temperature range},
  author={Benitez, R and Kan, WH and Gao, H and O'Neal, M and Proust, G and Srivastava, A and Radovic, M},
  journal={Acta Materialia},
  volume={189},
  pages={154--165},
  year={2020},
  publisher={Elsevier}
}

@article{garkas2010synthesis,
  title={Synthesis and characterization of Ti2AlC and Ti2AlN MAX phase coatings manufactured in an industrial-size coater},
  author={Garkas, Wagdi and Leyens, Christoph and Flores-Renteria, Arturo},
  journal={Advanced Materials Research},
  volume={89},
  pages={208--213},
  year={2010},
  publisher={Trans Tech Publ}
}

@article{HAFTANI201651,
title = {Studying the oxidation of Ti2AlC MAX phase in atmosphere: A review},
journal = {International Journal of Refractory Metals and Hard Materials},
volume = {61},
pages = {51-60},
year = {2016},
issn = {0263-4368},
doi = {https://doi.org/10.1016/j.ijrmhm.2016.07.006},
url = {https://www.sciencedirect.com/science/article/pii/S0263436816300786},
author = {Mohammad Haftani and Mina {Saeedi Heydari} and Hamid Reza Baharvandi and Naser Ehsani},
}

@article{carreras2017dynaphopy,
  title={DynaPhoPy: A code for extracting phonon quasiparticles from molecular dynamics simulations},
  author={Carreras, Abel and Togo, Atsushi and Tanaka, Isao},
  journal={Computer Physics Communications},
  volume={221},
  pages={221--234},
  year={2017},
  publisher={Elsevier}
}

@article{kresse2001vasp,
  title={VASP the Guide},
  author={Kresse, Georg},
  journal={https://cir.nii.ac.jp/crid/1574231875522562304},
  year={2001},
}

@misc{kresse2007vasp,
  title={VASP the guide. University of Vienna, Vienna},
  author={Kresse, G and Hafner, J},
  year={2007}
}

@article{kresse1996efficient,
  title={Efficient iterative schemes for ab initio total-energy calculations using a plane-wave basis set},
  author={Kresse, Georg and Furthm{\"u}ller, J{\"u}rgen},
  journal={Physical review B},
  volume={54},
  number={16},
  pages={11169},
  year={1996},
  publisher={APS}
}

\end{document}